# Frequency-switching Coherent Reception for Hardware-efficient High-baud-rate Optical Transmission Experiments

Hiroshi Yamazaki, *Member, IEEE*, Kohki Shibahara, *Member*, *IEEE*, Masanori Nakamura, *Member*, *IEEE*, Takayuki Kobayashi, *Member*, *IEEE*, Toshikazu Hashimoto, and Yutaka Miyamoto, *Member, IEEE*

***Abstract*— We investigate frequency-switching coherent reception (FSCR), a hardware-efficient technique for offline coherent optical signal characterization in laboratory environments, enabling bandwidth extension without costly and hardware-intensive parallel receiver architectures. By combining signal gating and local-oscillator frequency switching, FSCR captures multiple spectral slices sequentially using a single receiver front-end and oscilloscope. We experimentally demonstrate characterization of signals at symbol rates up to 288 GBaud beyond the receiver bandwidth limit.**

## I. INTRODUCTION

DRIVEN by exponential growth of data traffic, intensive research is underway to achieve per-wavelength throughput of 2 Tbps/λ or higher in coherent optical transmission systems [1-7]. Some experiments involved a technique called optical arbitrary waveform measurement (OAWM), in which a parallel receiver and frequency-comb local oscillator (LO) are used to effectively extend the receiver bandwidth by frequency (de-)multiplexing [4], [7], [8]. As the signal bandwidth rapidly increases, even high-end laboratory-use receiver components, such as balanced photodiodes (BPDs) and analog-to-digital converters (ADCs) with 100-GHz-class analog bandwidths, are becoming bandwidth bottlenecks, making such extension techniques increasingly important. However, the conventional OAWM poses a serious cost issue, as it requires a multiple of such high-end components.

In the field of space-division multiplexing (SDM), time-domain gating and mode switching have been used to avoid receiver parallelization for offline experiments in laboratories [9], [10]. By extending the concept to signal-to-LO-phase switching, a single-ADC offline coherent receiver has also been demonstrated [11]. Subsequently, by extending the concept to LO-frequency switching, a technique to extend the effective receiver bandwidth has also been demonstrated [12].

In this work, we investigate this technique, hereafter referred to as frequency-switching coherent reception (FSCR), for state-of-the-art high-speed signal characterization using modern high-bandwidth instruments. Using a single 90-GHz coherent receiver frontend (FE) with four BPDs and a single 110-GHz digital storage oscilloscope (DSO) with four ADCs, we characterize dual-polarization quadrature amplitude modulation (DP-QAM) signals with symbol rates up to 288 GBaud and net bit rates (NBRs) up to 2.94 Tbps, demonstrating applicability of this technique to current multi-Tbps/λ-class coherent signal characterization.

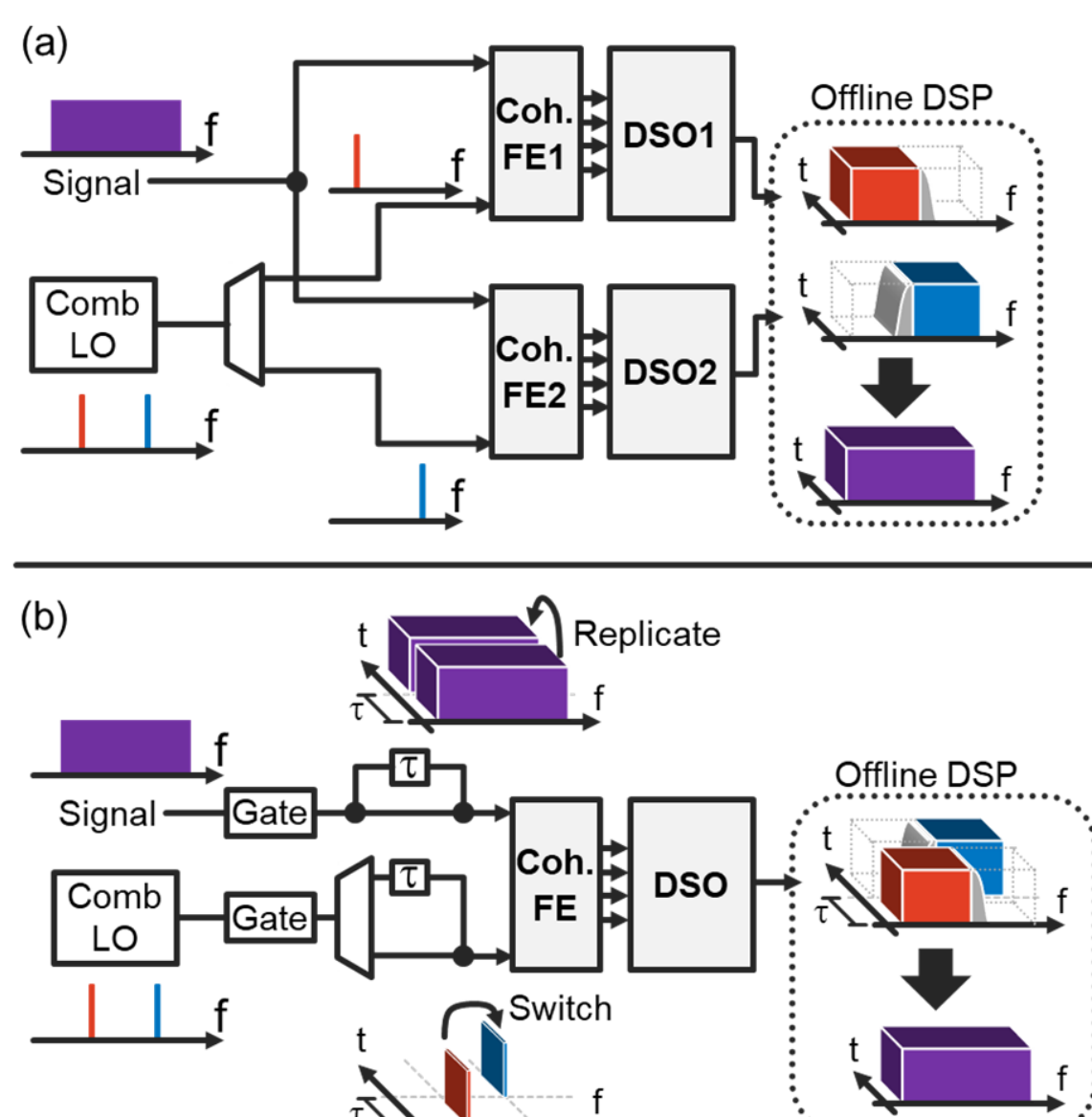

**Fig. 1.** Principles of (a) parallel OAWM and (b) FSCR.

## II. PRINCIPLE

Figures. 1(a) and (b) show the principles of parallel OAWM and FSCR, respectively, both for receiving signals with bandwidths up to nearly twice that of each FE and DSO. In the parallel OAWM, a signal under test is bifurcated (or frequency demultiplexed) and sent to the two FEs, where they are mixed with two different lines of the comb LO represented in red and blue, respectively. The down-converted RF signals output from the two FEs correspond to two different frequency slices of the signal. They are then captured by two separate DSOs and processed offline where they are stitched to each other in the frequency domain to reconstruct the whole signal. In the FSCR, the signal under test is gated and sent to a time-domain multiplexer with a delay τ, which is slightly larger than the opening duration of the gate, so that a replica of the gated signal is added closely after the original. The comb LO is also gated synchronously to the signal before being frequency demultiplexed, and the two gated frequency lines are delayed by τ relative to each other before being added, leading to LO-frequency switching. The duplicated gated signal and frequency-switching LO are then mixed in the FE, generating two consecutive RF bursts, each corresponding to one of the two frequency slices of the signal. Finally, they are captured by the DSO and processed offline where they are time demultiplexed and frequency stitched to reconstruct the whole signal. For offline experiments, time gating is generally acceptable since only finite-duration signal segments are analyzed in any case. FSCR requires roughly double the length of the waveform to be captured by the DSO, but this does not pose a significant problem; in most cases, the memory length of the DSO is sufficient for coherent signal characterization at Tbps/λ-class data rates, as detailed later. Thus, FSCR, while inherently limited to offline experiments, provides essentially the same functionality as parallel OAWM within this offline measurement framework, requiring only a single FE and single DSO.

## III. Experiment

Figure. 2 shows the experimental setup for the dual-polarization (DP) QAM signal characterization using FSCR, along with the measured optical spectra, which are described later. We generated the DP-QAM signals with symbol rates up to 288 GBaud by using a colorless optical-domain bandwidth extension technique [13]. The transmitter consisted of an offline PC, a four-channel arbitrary waveform generator (AWG, Keysight M8199B) operating at 252 GS/s, indium-phosphide delayed dual IQ modulator (D2IQM) [6], external cavity laser (ECL) oscillating at a carrier frequency, $f_c$, of 193.4 THz, straight phase modulator (PM), and wavelength selecting switch (WSS). The driver amplifier integrated in the D2IQM had a steep cutoff at around 72 GHz. Accordingly, we input a 72-GHz carrier-suppressed-return-to-zero (CSRZ) pulse train to the D2IQM by driving the PM at 36 GHz and extracting comb lines at ±72 GHz from fc with the WSS, as shown by the left optical spectra in Fig. 2. By using an offline digital spectral weaver (DSW) with this setup, we can generate arbitrary optical signals with bandwidths on each side of fc up to 144 GHz, which is twice that of the drivers. The CSRZ frequency of 72 GHz deviates significantly from the design condition of 56 GHz for the D2IQM used for this experiment, but it is still acceptable as the DSW inherently compensates for this mismatch as well as other hardware imperfections. Using the D2IQM, we generated 16-level QAM (16QAM) signals with symbol rates ranging from 144 to 288 GBaud and probabilistically-constellation-shaped 144-level QAM (PCS-144QAM) signals at 288 GBaud with varying entropy. We set the signal length and pilot overhead to ~$3\times10^5$ symbols and 0.79%, respectively. After the D2IQM, we used an external polarization-division-multiplexing emulator (PDME) with a relative delay of 65.7 ns and optical bandpass filter before the transmission line, which was switchable between 0 and 80 km of standard single-mode fiber (SSMF). On the receiver side, we used a gate switch driven by a 20-kHz rectangular wave with a duty cycle of 49% followed by a time-domain multiplexer with a relative delay of 25 µs to gate and replicate the signal. As the LO source, we used another free-running ECL oscillating at around fc. We split the LO light to the continuous-wave (CW) and frequency-switching (FS) paths. In the FS path, we used a PM driven at 40 GHz, gate switch synchronized to the one in the signal line, WSS extracting two comb lines at ±80 GHz from the $f_c$ separately to two output ports, 25-µs delay line applied to the +80-GHz line, and combiner. The corresponding optical spectra at the WSS input and outputs are shown on the right in Fig. 2. The CW path was implemented to obtain reference data, and an optical switch for selecting between the CW LO and the FS LO was placed before a coherent FE with an opto-electronic 3-dB bandwidth of around 90 GHz. The FE was followed by a 256-GS/s 110-GHz four-channel DSO (Keysight UXR1104A), which was triggered to capture the two consecutive bursts to be analyzed offline. Assuming an information rate of 10 bits/symbol at 288 GBaud, the burst length of 25 µs corresponds to $7.2\times10^7$ bits ($7.2\times10^6$ symbols), which is sufficient for analyses of signals with target bit error rates (BERs) of an order of $10^{-5}$ after decoding of soft-decision forward-error-correction (SD-FEC) codes. The memory length of the DSO is around 780 µs (200 Mpts), which is more than enough for the condition described above.

Figure. 3 illustrates the processing flow of the offline receiver digital signal processing (DSP). After correction of the FE, the two consecutive bursts corresponding to lower and higher-frequency (LF and HF) slices, respectively, were extracted, and the HF burst was sent to the chromatic dispersion compensator (CDC) for the 5-km (25-µs) delay line. The process after that was basically the same as that for parallel OAWM; the LF and HF bursts were frequency-shifted by −80 and +80 GHz, respectively, before being stitched to each other by matching amplitudes and phases of the overlapping parts. The process up to this point is completely blind, using only captured waveform data. We then used the same processing blocks and assumed the same FEC configuration as those used in previous studies where a frequency-domain adaptive equalizer (AEQ) with a pilot-based digital phase-lock loop (DPLL) was applied before calculating the signal-to-noise ratios (SNRs) from the received constellation, achievable information rate (AIR) from the normalized generalized mutual information (NGMI), and NBR from the required code rate for error-free decoding after the FEC [3], [6]. Figure. 3 also shows, as an example, the time-domain waveform of the extracted bursts and corresponding frequency-shifted spectra for a 288-GBaud signal.

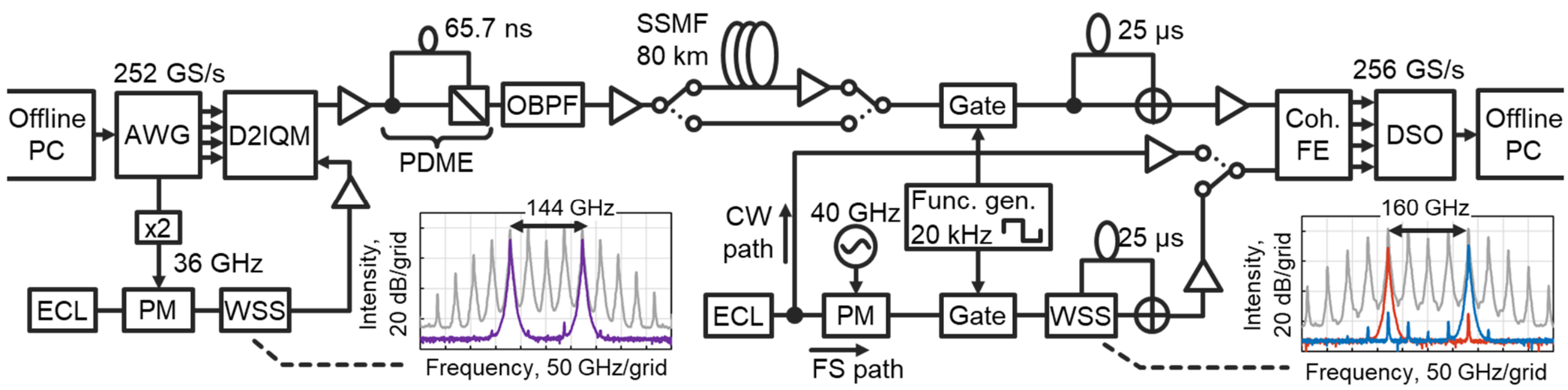

**Fig. 2.** Experimental setup.

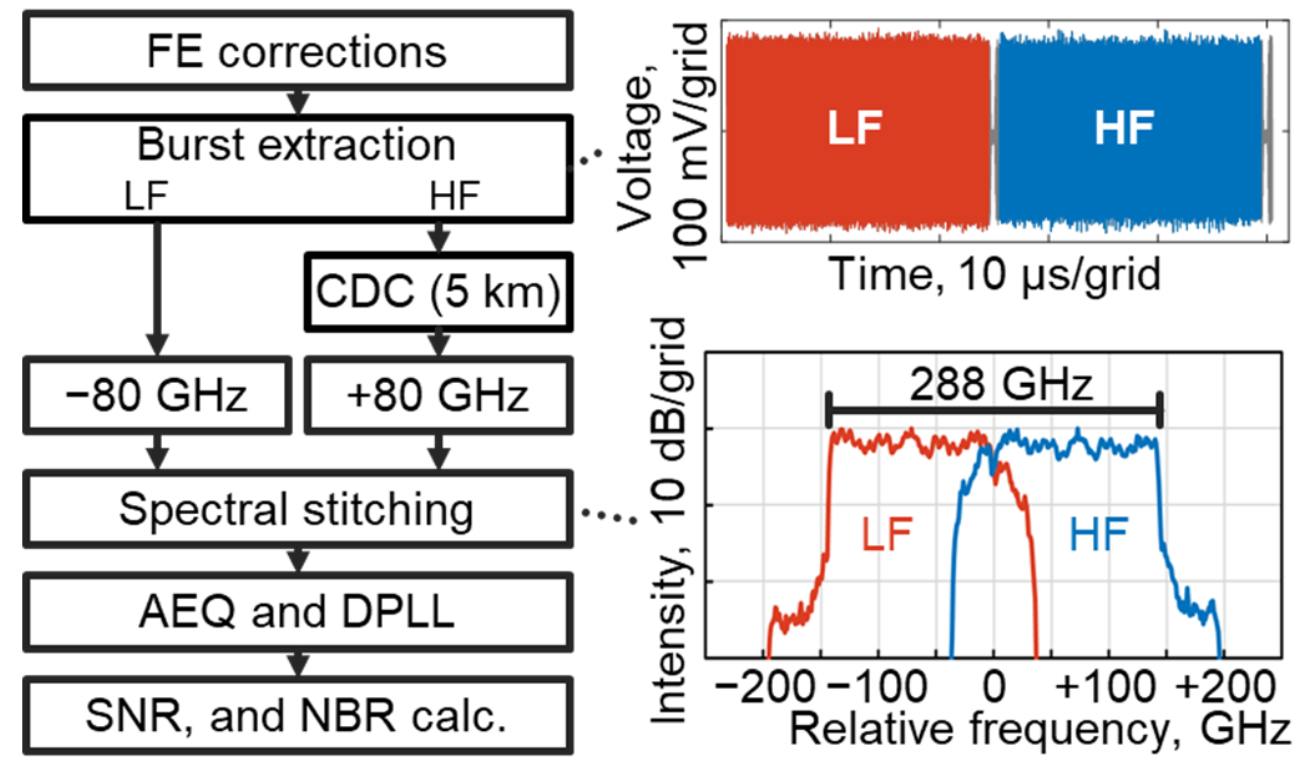

**Fig. 3.** Processing flow of offline receiver DSP for FSCR, time-domain waveform of the extracted bursts, and corresponding frequency-shifted spectra for a 288-Gbaud signal.

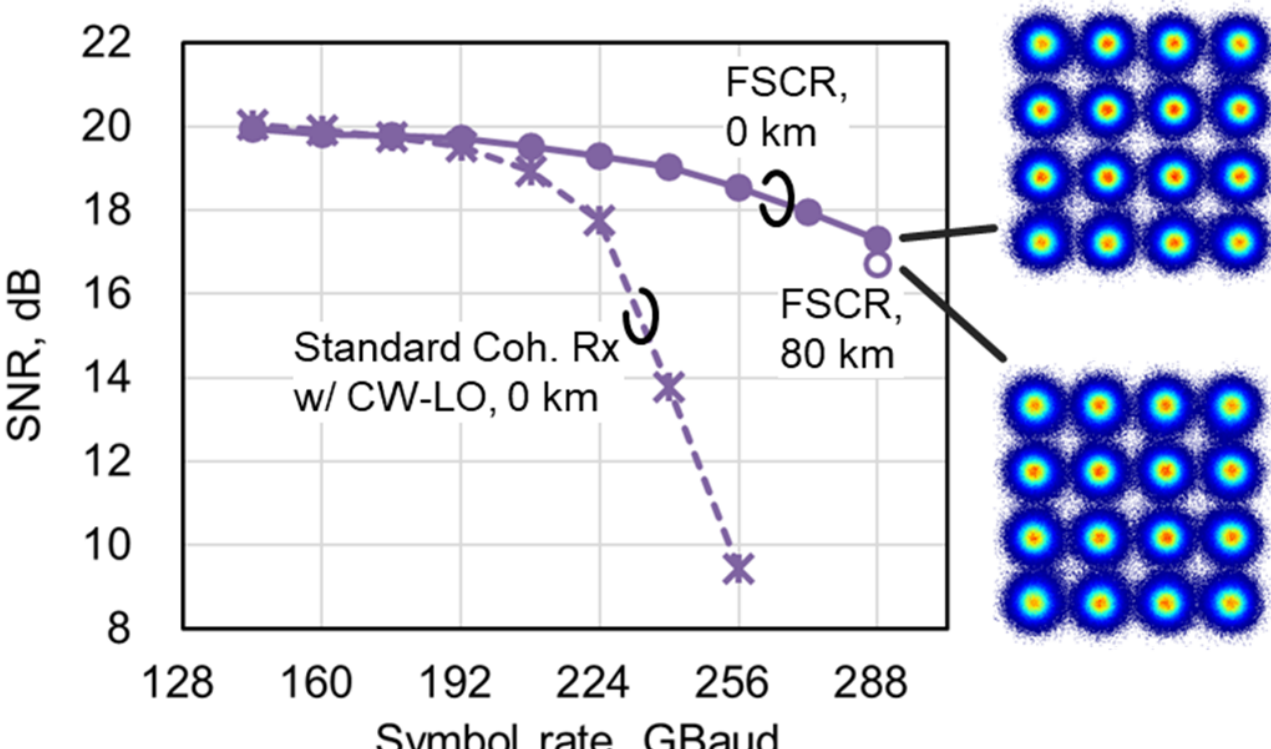

**Fig. 4.** SNR vs symbol rate of 16QAM signals.

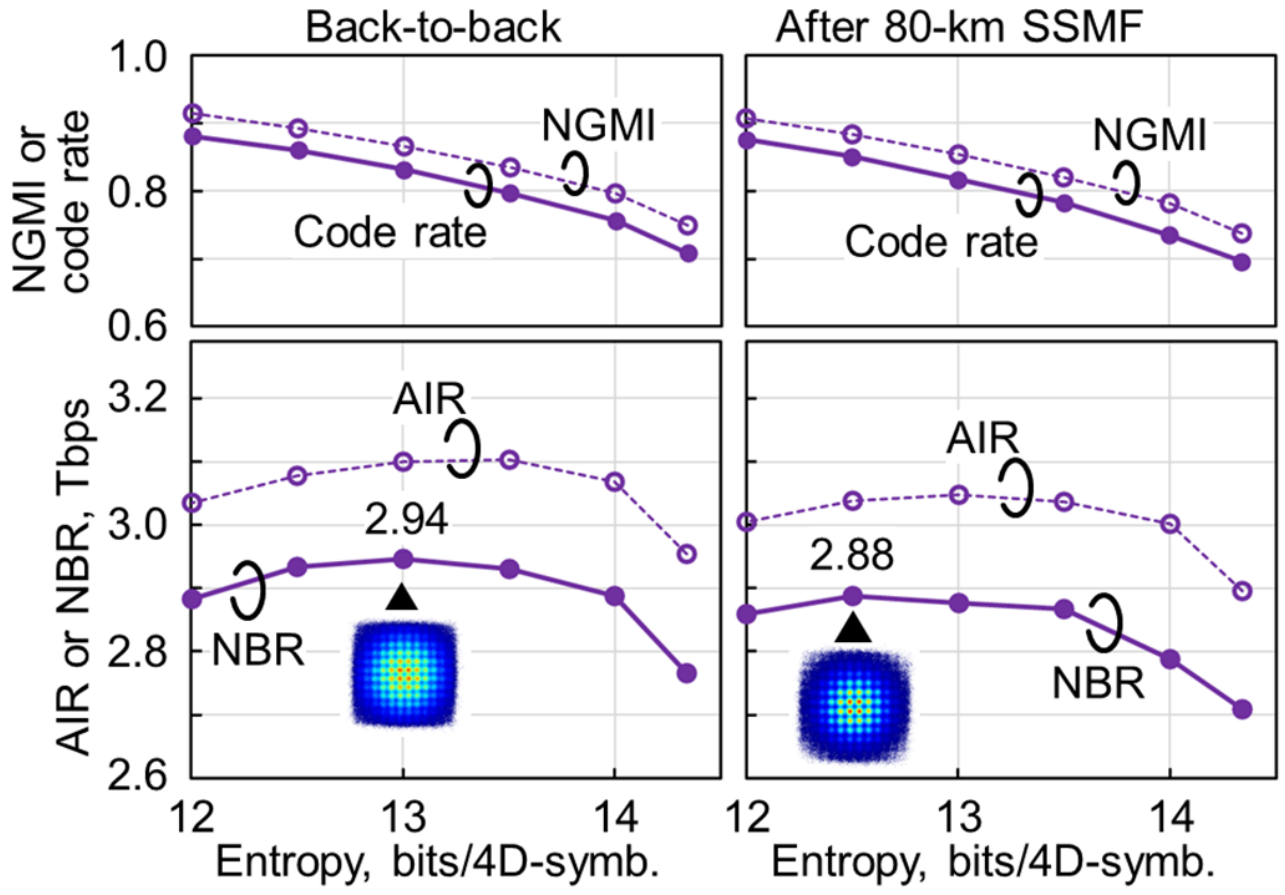

**Fig. 5.** NGMI, code rate, AIR, and NBR vs entropy of PCS-144QAM signals.

Figure. 4 shows the measured SNRs of the 16QAM signals as a function of symbol rate for both standard coherent reception with the CW LO and FSCR with the FS LO. With standard coherent reception, the SNR rapidly decreased as the symbol rate exceeded 200 GBaud due to the bandwidth limitation of the coherent FE and DSO. Using FSCR, however, we successfully received the DP-16QAM signals over the entire tested range of symbol rates, with SNRs of 17.3 and 16.7 dB at 288 GBaud for 0- and 80-km transmissions, respectively. Figure. 5 shows the NGMI, optimized code rate, AIR, and the NBR obtained by varying entropy, $H$, for each four-dimensional symbol of the PCS-144QAM signals. We calculated the AIR and NBR in Tbps by $C=\{H-(1-R)\times16\}/1.0079\times0.288$, where $C$ is AIR or NBR and $R$ is the NGMR (for AIR) or code rate (for NBR). The maximum NBRs were 2.94 Tbps for back-to-back and 2.88 Tbps after 80-km SSMF transmission.

## IV. Conclusion

FSCR is a promising technique to scale receiver bandwidth in offline optical signal characterization without resorting to parallelization using multiple sets of costly measurement hardware. This technique will make characterization of multi-Tbps/λ-class optical signals more accessible and accelerate research and development of next-generation digital coherent components and systems.